\documentclass[a4paper,11pt]{article}
\usepackage{pos}
\usepackage{physics}
\usepackage{cancel}
\usepackage{feynmp}
\usepackage{graphicx}
\usepackage{caption}
\usepackage{subcaption}
\usepackage{svg}
\usepackage{multirow}
\usepackage{multicol}
\usepackage{makecell}
\usepackage{footnote}
\usepackage{mathtools}
\usepackage{orcidlink}
\usepackage{booktabs}
\usepackage{enumitem}
\def\D{\mathrm{D}}

\DeclareRobustCommand{\LoopIn}{\textsc{LoopIn}}

\title{Symbolic syzygy-constrained reduction rules for Feynman integrals and the LoopIn framework}

\author*{Sid Smith}

\affiliation{Dipartimento di Fisica e Astronomia, Universita di Padova, Via Marzolo 8, 35131 Padova, Italy}

\affiliation{INFN, Sezione di Padova,
Via Marzolo 8, I-35131 Padova, Italy.}

\affiliation{Higgs Centre for Theoretical Physics, University of Edinburgh, James Clerk Maxwell Building,Peter Guthrie Tait Road, Edinburgh, EH9 3FD, United Kingdom}

\emailAdd{sid.smith@ed.ac.uk}

\abstract{We present a new algorithm for integration-by-parts (IBP) reduction of Feynman integrals with high powers of numerators or propagators, a demanding computational step in evaluating multi-loop scattering amplitudes. The algorithm allows us to avoid a large intermediate system of equations and instead focus on applying direct reduction rules to the integrals. We demonstrate the application of our algorithm with some highly non-trivial examples, namely rank-20 integrals for the double box with an external mass and the massless pentabox. We also achieve much faster IBP reduction for an example of scattering amplitudes for spinning black hole binary systems. Finally, we present LoopIn, a modular framework for automating multi-loop calculations, where the IBP techniques described here can be interfaced.}

\FullConference{17th International Symposium on Radiative Corrections: Applications of Quantum Field Theory to Phenomenology (RADCOR2025)\\
5-10 October 2025\\
Puri, India\\}

\begin{document}

\maketitle

\begin{fmffile}{feyndiagrams}

\section{Introduction}

Integration-by-parts reduction \cite{Chetyrkin:1981qh,Laporta:2000dsw} of multi-loop Feynman integrals is one of the most significant computational bottlenecks in modern scattering amplitude calculations. Recent significant developments include finite field reconstruction \cite{Kant:2013vta,vonManteuffel:2014ixa,Peraro:2016wsq,Abreu:2018zmy,Klappert:2019emp,Peraro:2019svx,Laurentis:2019bjh,DeLaurentis:2022otd,Magerya:2022hvj,Belitsky:2023qho,Chawdhry:2023yyx,Liu:2023cgs,Maier:2024djk}, syzygy equation constraints \cite{Tarasov:2004ks,Smirnov:2005ky,Smirnov:2006tz,Smirnov:2006wh,Gerdt_2006,Lee:2008tj,Barakat:2022qlc,Bendle:2019csk,Ita:2015tya,Larsen:2015ped,Gluza:2010ws,Wu:2023upw,Wu:2025aeg} and improved seeding methods \cite{Driesse:2024xad, Guan:2024byi, Bern:2024adl, Lange:2025fba,vonHippel:2025okr,Song:2025pwy,Zeng:2025xbh,Smirnov:2025prc,Brunello:2025gpf,Bern:2025wyd,Driesse:2026qiz}. Despite these advancements, the significant computational challenge remains in solving large linear systems of IBP equations. These challenges are bypassed when one produces symbolic reduction rules that directly reduce arbitrary target integrals to lower-complexity integrals. This approach has been explored in Refs.~\cite{Ruijl:2017cxj,Kosower:2018obg,Guan:2023avw,Feng:2025leo} and is implemented in LiteRed \cite{Lee:2012cn,Lee:2013mka}, however the use of symbolic reduction rules has found limited applications in QFT calculations relative to the standard approach. Specifically, the use of symbolic reduction rules for multi-scale Feynman integrals beyond one loop is largely unexplored territory.

In this paper, we present a novel algorithm to generate symbolic reduction rules that can be applied to a specific set of target integrals. The algorithm combines the methods of syzygy constraints, smart seeding, and judicious reshuffling of the identities, in order to generate these rules.
A main feature of our approach is that we aim to keep the explicit dependence of the reduction rules on the propagator/numerator powers $n_{i}$ minimal, by performing Gaussian elimination at the level of IBP operators in a way that is agnostic to the seed integrals used.\footnote{Similar operator-level reshuffling, though in the Laporta approach without syzygy equations, was silently added in the version 6.5 of FIRE \cite{Smirnov:2023yhb} and improved in the version 7 \cite{Bern:2024adl,Smirnov:2025prc}. The aim there is simplifying the IBP operators before normal seeding and linear solving, rather than directly finding reduction rules.} For a subset of Feynman integrals that cannot be reduced by the above rules, we additionally solve a linear system of equations formed by a small number of seed integrals in the neighborhood of the targeted integrals, to produce extra symbolic reduction rules to complete the IBP reduction. \footnote{In practice, this subset of integrals may depend on a number of factors, such as the monomial ordering chosen in the syzygy computation, or the ordering of the indices used to define the weight of an integral. There could potentially exist an optimal choice for each specific case, which is a problem that would require further exploration.}
We are able to use these rules to reduce integrals with multiple powers of different ISPs for highly nontrivial multi-scale integral topologies such as the two-loop pentabox, as we show in the examples.

\section{Integration-by-Parts: An Overview \label{sec:IBP}}

\paragraph{Integration-by-Parts Identities}

We denote an $L$-loop Feynman integral using the following notation
\begin{equation}
    F(n_{1},\dots,n_{N}) = \int\left(\prod_{a=1}^{L}\dd^{\D}\ell_{a}\right)\frac{1}{\rho_{1}^{n_{1}}\dots\rho_{N}^{n_{N}}}\,, \label{eq:family}
\end{equation}
where $\rho_{i}$ are the propagators, given in terms of scalar products involving the loop momenta $\ell_{a}$ and a set of external momenta $p_{i}$, $i=1,\dots,E$. We define the \textit{rank} $r$ and number of \textit{dots} d as follows:
\begin{equation}
    r(\vec n) = -\sum_{n_{i}<0}n_{i}, \quad d(\vec n) = \sum_{n_{i}>1}(n_{i}-1)\,.
\end{equation}
The aim is to express any arbitrary Feynman integral in terms of it's master integrals $J_{i}$:
\begin{equation}\label{eq:master_decomposition}
    F(\vec n) = \sum_{i}c_{i}J_{i}\,.
\end{equation}
We can find these coefficients $c_{i}$, by starting with the IBP identity
\begin{equation}
    0 = \int\left(\prod_{a=1}^{L}\dd^{\D}\ell_{a}\right)\pdv{}{\ell_{b}^{\mu}}\left(\frac{q_{\alpha}^{\mu}}{\rho_{1}^{n_{1}}\dots\rho_{N}^{n_{N}}}\right)\,, \label{eq:ibp}
\end{equation}
where $q_{\alpha}^{\mu}\in\left\{\ell_{1},\dots,\ell_{L},p_{1},\dots,p_{E}\right\}$ and $b=1,\dots,L$. By choosing a specific combination of $(b,\alpha,\vec n)$ values, we can generate a system of equations. By solving this linear system of equations, we can find exactly the coefficients $c_{i}$ in Eq.~\eqref{eq:master_decomposition}.

\paragraph{Sectors}

In this work, we describe the \textit{sector} $s$ of an integral by the positive entries of $\vec n$:
\begin{equation}
    s(\vec n) = \vec n\eval_{-\text{ve}\rightarrow0}
\end{equation}
For example, the integral $F(2,1,1,-1,-2)$ has sector $s(2,1,1,-1,-1) = (2,1,1,0,0)$.  This is different to the standard definition of a sector, which is only concerned with which propagators have a positive power, not the specific value they take.

Given two integrals $F(\vec n_{1})$ and $F(\vec n_{2})$, we define $F(\vec n_{1})$ to belong to a \textit{subsector} of $F(\vec n_{2})$ iff
\begin{equation}
    s(\vec n_{1})_{i}\geq s(\vec n_{2})_{i}\quad\forall\,i
\end{equation}
This definition is convenient when syzygy constraints are introduced, as it is consistent with the idea that IBP identities in a given sector will relate integrals within that sector and subsectors of it.

\paragraph{Syzygy Constraints}

We can constrain the types of equations we generate in our system by enforcing constraints when building the identities. We begin by extending the IBP identity from Eq.~\eqref{eq:ibp}
\begin{equation}
    0 = \int\left(\prod_{a=1}^{L}\dd^{\D}\ell_{a}\right)\pdv{}{\ell_{b}^{\mu}}\left(\frac{P_{b\alpha}(\rho)q_{\alpha}^{\mu}}{\rho_{1}^{n_{1}}\dots\rho_{N}^{n_{N}}}\right)\,,
\end{equation}
where $P_{a\alpha}(\rho)$ are polynomials in the propagators. This is equivalent to taking a linear combination of the previous identities with different seed choices $\vec n$. We can then enforce the constraints
\begin{equation}
    P_{b\alpha}(\rho)q_{\alpha}^{\mu}\pdv{}{\ell_{b}^{\mu}}\rho_{i} = f_{i}(\rho)\rho_{i}, \quad \forall\, i\in\sigma\,,
\end{equation}
on a subset $\sigma$ of the propagators. Typically $\sigma$ is chosen to be all the propagators that appear as a denominator in the given sector $\sigma = \{i\,|\,s(\vec n)_{i}>0\}\,$. 

These constraints give control over the sectors of the integrals that appear in the resulting equations. For example if one of your seed integrals has $n_{i}=2$, you can expect only integrals with $n_{i}\leq2$ in the resulting equation. This significantly reduces the number of variables appearing in the final system of equations.

The constraint here can be seen as a \emph{syzygy equation} in the language of computational algebraic geometry (see \cite{Smith:2025xes} for details). We use Singular \cite{DGPS} to find the solutions to this, which give us choices we can make for the polynomials $P_{b\alpha}(\rho)$ that result in the desired identities.

\section{Reduction Rule Algorithm \label{sec:algorithm}}

In this section we describe our algorithm for reducing integrals of arbitrarily high complexity. This is split into two natural stages

\begin{enumerate}
    \item Generating Reduction Rules
    \item Applying Reduction Rules
\end{enumerate}

We use the following \textit{weight} function, which is vector-valued and interpreted lexicographically to order integrals,
\begin{equation}
    W(\vec n) = \left(\sum_{n_{i}>0}1,\, d(\vec n),\, r(\vec n), \, -\mathcal{O}(\abs{\vec n})\right)\,.
\end{equation}
This weight is determined by the number of positive indices, the number of dots, the rank and finally some arbitrary ordering function $\mathcal{O}$. A weighting function is crucial to have a notion of reduction rules.

\subsection{Generating Reduction Rules}

\begin{figure}[h!]
    \centering
    \includegraphics[scale=0.5]{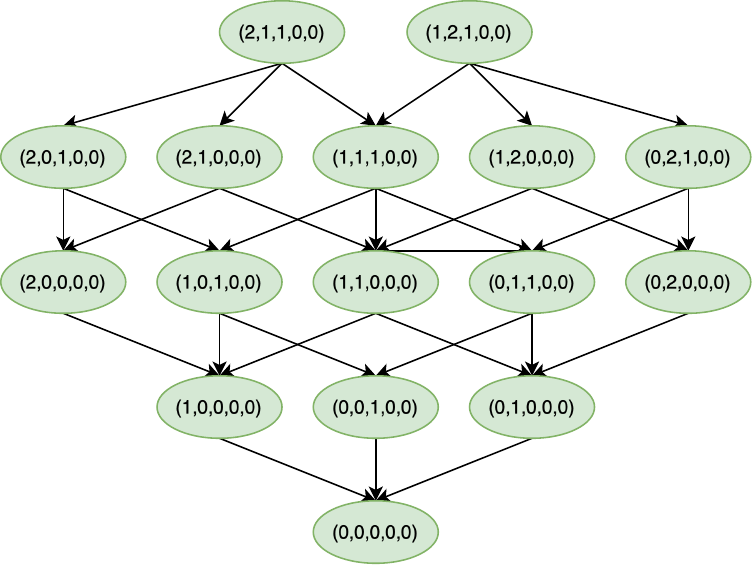}
    \caption{Tower of Sectors one must consider when the top sectors are $(2,1,1,0,0)$ and $(1,2,1,0,0)$. This drawing is schematic and the horizontal levels here do not correlate with the weighting described later. The relevant information is contained within the arrows, denoting the subsector inheritance hierarchy.\label{fig:sectors}}
\end{figure}

The algorithm aims to first generate reduction rules for each individual sector relevant to the specific target integrals. For example, the following target integrals translate to the sectors
\begin{equation}
    \begin{aligned}
    \{F(2,1,1,-4,-4),F(1,2,1,0,-7),F(1,1,1,-6,-4),F(1,1,1,-11,0)\}\\
    \implies \{(2,1,1,0,0),(1,2,1,0,0),(1,1,1,0,0),(1,1,1,0,0)\}\,.
    \end{aligned}
\end{equation}
From this we can remove duplicates and subsectors to obtain the set of \textit{top sectors}
\begin{equation}
    \{(2,1,1,0,0),(1,2,1,0,0)\}\,.
\end{equation}
We can then determine the tower of sub-sectors, as shown in Fig. \ref{fig:sectors}. This organisation of sectors is similarly done in NeatIBP \cite{Wu:2023upw,Wu:2025aeg}. Note also that one can remove vanishing sectors from this tower.

\begin{figure}[h!]
    \centering
    \includegraphics[scale=0.5]{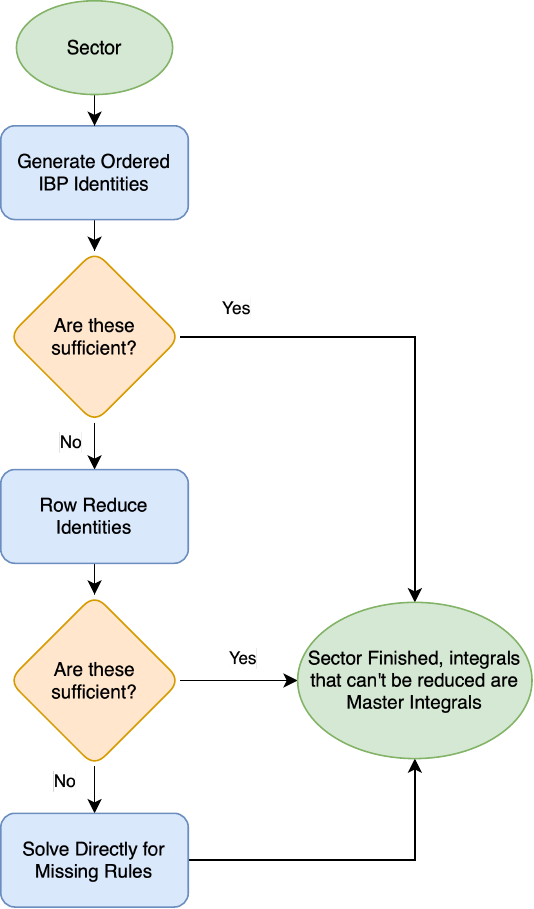}
    \caption{A flow chart describing the algorithm for generating reduction rules\label{fig:flowchart}}
\end{figure}

For each sector, the flow chart in Fig.~\ref{fig:flowchart} describes the algorithm for finding reduction rules. Each step is summarised below.

\paragraph{Generating Ordered Identities}

In each sector $\vec m$ from the tower, we solve the syzygy equations with constraints on the subset $\sigma = \{i\,|\,m_{i}>0\}$ of propagators
\begin{equation}
    P_{b\alpha}(\rho)q_{\alpha}^{\mu}\pdv{}{\ell_{b}^{\mu}}\rho_{i} = f_{i}(\rho)\rho_{i}, \quad \forall\, i\in\sigma\,.
\end{equation}
In practice, this can be done for each unique subset $\sigma$ that appears on the tower, instead of the full set of sectors. We can write the resulting identities in the form
\begin{equation}\label{eq:ibpIdentities}
    0 = \sum_{i}(\alpha_{i}+\vec{\beta}_{i}\cdot\vec{n})F(\vec{n}+\vec{\gamma}_{i})\,.
\end{equation}
Since we are only interested in reducing integrals in the sector $\vec m$, namely integrals that have $n_{i}=m_{i}$ for $i\in\sigma$, we are free to input the seed integral given by
\begin{equation}
    n_{i} = \begin{cases}
        m_{i},& i\in\sigma\,,\\
        \eta_{i},& i\notin\sigma\,.
    \end{cases}\\
\end{equation}
where $m_{i}$ are the fixed integers for the sector, and $\eta_{i}$ are still free to be
anything. This is equivalent to shifting
$\alpha_{i}\rightarrow\alpha_{i}+\vec\beta_{i}\cdot\vec m$ and
shortening $\vec\beta_{i}$ to only contain the non-sector
indices. This constitutes the main motivation for using syzygy
constraints in our reduction rules, as it allows us to reduce the number of free indices from $N\to N-\abs{\sigma}$.

We can obtain initial reduction rules from the identities by identifying the highest weight shift vector, using the following sector-specific weight function
\begin{equation}
    w(\vec{\gamma}) = (\vec{\gamma}\cdot\vec{\xi},-\vec{\gamma}\cdot\vec{\theta},-\mathcal{O}(\vec{\gamma}))\,,\quad \vec{\xi} = \vec{m}\eval_{+\text{ve}\rightarrow1}, \quad \theta_{i} = 1-\xi_{i}\,,
\end{equation}
This orders shifts $\vec \gamma_i$ in Eq.~\eqref{eq:ibpIdentities}  according to whether they result in subsectors, then by rank and finally by the same ordering function.

\paragraph{Row Reduce Identities}

If the reduction rules generated in the previous step are insufficient to reduce all integrals on the sector, we move onto this step. By using seed integrals that correspond to small perturbations around the integrals we are unable to reduce with the existing rules, we produce additional identities. Gathering up and ordering the $M$ shift vectors present, we can represent the $I$ identities as a tensor product
\begin{equation}
    \begin{pmatrix}
        \alpha_{11}&\vec\beta_{11}&\cdots&\alpha_{1M}&\vec\beta_{1M}\\
        \vdots&\vdots&\vdots&\vdots&\vdots\\
        \alpha_{I1}&\vec\beta_{I1}&\cdots&\alpha_{IM}&\vec\beta_{IM}
    \end{pmatrix}\otimes\begin{pmatrix}
        \vec\gamma_{1}\\
        \vec\gamma_{2}\\
        \vdots\\
        \vec\gamma_{M}
    \end{pmatrix}\,. \label{eq:matrixToRowReduce}
\end{equation}
The notation used here with the tensor product is a short hand for the identities in Eq.~\ref{eq:ibpIdentities}, each row in the first matrix corresponds to an identity
\begin{equation}
    \begin{pmatrix}
        \alpha_{k1}&\vec\beta_{k1}&\cdots&\alpha_{kM}&\vec\beta_{kM}
    \end{pmatrix}\otimes\begin{pmatrix}
        \vec\gamma_{1}\\
        \vec\gamma_{2}\\
        \vdots\\
        \vec\gamma_{M}
    \end{pmatrix}[\vec n] = \sum_{i=1}^{M}(\alpha_{ki}+\vec\beta_{ki}\cdot\vec n)F(\vec n+\vec\gamma_{i})\,.
\end{equation}
After performing a row reduction to the matrix of coefficients, we are left with new identities, and some of these may give us additional reduction rules that we can add to our set.
The row reduction is done using FiniteFlow, and the rows corresponding to relevant reduction rules are reconstructed. 

After this step is done once, there may be a different set of integrals we are unable to reduce, for which we can use a different set of seed integrals as small perturbations. Therefore, this step is performed iteratively until we cannot find any more relevant rules.

\paragraph{Solving Directly for Missing Rules}

Finally, if there still exist integrals which the current list cannot reduce, we attempt to find symbolic reduction rules for these in terms of the indices $n_{i}$. We do this by generating a small system of equations around the fixed point specified by these integrals, i.e.\ using seed integrals in the neighborhood of the targeted integrals, keeping the analytic dependence on the indices $n_{i}$. We then solve analytically using FiniteFlow for the reduction rule associated to the integral. This generally produces results with non-polynomial rational dependence on $n_{i}$, unlike the operator-level row-reduction with Eq.~\eqref{eq:matrixToRowReduce}. After this step, if there are any integrals remaining on the sector that we are unable to reduce, these are interpreted as master integrals.




\subsection{Applying Reduction Rules}

We apply the reduction rules through backward substitution. This is done by creating an equation that acts as a reduction rule for every target integral in our list, we then create a similar equation for all the integrals that appear in these created equations. This is done iteratively until we have a system containing, by definition, no redundant equations.

Gaussian eliminination comes in two steps, forward elimination to put the matrix into a row echelon form, and backward substitution, which brings it to a \emph{reduced} row echelon form. This is depicted below for a simple example below, where the last column of the matrix corresponds to the only "master integral",
\begin{equation}\label{eq:backsub}
    \begin{pmatrix}
        \#&\#&\#&0\\
        \#&\#&0&\#\\
        \#&0&0&\#\\
        0&\#&\#&\#\\
        0&0&\#&0
    \end{pmatrix}\xrightarrow{\text{forward elimination}}\begin{pmatrix}
        \#&\#&\#&0\\
        0&\#&\#&\#\\
        0&0&\#&\#\\
        0&0&0&\#\\
        0&0&0&0
    \end{pmatrix}\xrightarrow{\text{backward substitution}}\begin{pmatrix}
        \#&0&0&\#\\
        0&\#&0&\#\\
        0&0&\#&\#\\
        0&0&0&\#\\
        0&0&0&0
    \end{pmatrix}\,.
\end{equation}
In our case, the system we generate is already in row echelon form by definition, so we just need to perform back substitution, which is typically a far faster step than forward elimination for numerical (finite-field) Gaussian elimination.

\section{Examples \label{sec:ex}}

In this section we apply our algorithm to some highly non-trivial examples. The implementation of our algorithm was written in Mathematica and used FiniteFlow \cite{Peraro:2019svx} and Singular \cite{DGPS}, all computations were performed on a laptop.

In each case we find that the number of equations needed is significantly less than what would be needed in the standard approach. This is not a direct comparison as the equations that are considered here are much more dense. The sample time for applying these reduction rules using backward substitution at a single numerical point modulo some prime number is very small compared with the equation generation times presented in Table~\ref{tab:tab2} and Table~\ref{tab:tab3}. However it is well known that the IBP coefficients for the reduction of these complex integrals are highly non-trivial, and therefore rational function reconstruction would still be a bottleneck here.

\subsection{Double Box with External Mass}

\begin{figure}[h!]
    \centering
    \begin{fmfgraph*}(200,100)
        \fmfforce{(0,0)}{i1}
        \fmfforce{(0,h)}{i2}
        \fmfforce{(w,0)}{i3}
        \fmfforce{(w,h)}{i4}
        \fmfforce{(0.2w,0.2h)}{v1}
        \fmfforce{(0.2w,0.8h)}{v2}
        \fmfforce{(0.5w,0.2h)}{v3}
        \fmfforce{(0.5w,0.8h)}{v4}
        \fmfforce{(0.8w,0.2h)}{v5}
        \fmfforce{(0.8w,0.8h)}{v6}
        \fmf{fermion,label=$p_{1}$}{i1,v1}
        \fmf{fermion,label=$p_{2}$}{i2,v2}
        \fmf{fermion,label=$p_{3}$}{i4,v6}
        \fmf{dbl_plain}{i3,v5}
        \fmf{fermion,label=$\ell_{1}$}{v1,v3}
        \fmf{fermion,label=$\ell_{2}$}{v3,v5}
        \fmf{plain}{v2,v4,v6}
        \fmf{plain}{v1,v2}
        \fmf{plain}{v3,v4}
        \fmf{plain}{v5,v6}
    \end{fmfgraph*}
    \caption{Two Loop Box\label{fig:doublebox}}
\end{figure}
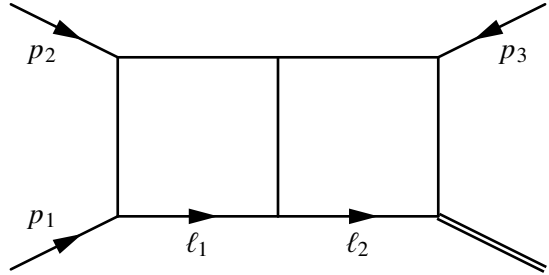

The first example is the double box with an external mass, as shown in Fig.~\ref{fig:doublebox}. We consider the following list of target integrals
\begin{equation}
    \{F(1,1,1,1,1,1,1,-10,-10),F(1,2,1,1,1,1,1,-6,-6),F(1,1,1,1,1,1,1,-2,-15)\}\,,
\end{equation}
which go up to rank $20$ with one dot. We provide statistics for the equation generation for this set of integrals in Table~\ref{tab:tab2}.
\begin{table}[h!]
\centering
    \begin{tabular}{|c|c|c|c|c|}
        \hline
        Cut&Time Taken&Number of Equations&Number of Masters\\
        \hline
        \{5,7\}&284s&18971&14\\
        \hline
        \{1,4,7\}&60s&8120&4\\
        \hline
        \{3,6,7\}&178s&12287&8\\
        \hline
        \{4,6,7\}&57s&2643&4\\
        \hline
        \{1,3,4,6\}&26s&1153&3\\
        \hline
        \{1,3,5,6\}&35s&2031&5\\
        \hline
    \end{tabular}
    \caption{Table showing the time taken for generating the system of reduction rules \textit{analytically} on a set of spanning cuts, as well as the number of rules applied (equations) and master integrals. The sample time for applying these reduction rules using backwards substitution at a single numerical point modulo some prime number is very small compared to the times presented here.\label{tab:tab2}}
\end{table}

\subsection{Massless Pentabox}

\begin{figure}[h!]
    \centering
    \begin{fmfgraph*}(200,100)
        \fmfforce{(0,0)}{i1}
        \fmfforce{(0,h)}{i2}
        \fmfforce{(0.9w,0)}{i3}
        \fmfforce{(0.9w,h)}{i4}
        \fmfforce{(w,0.5h)}{i5}
        \fmfforce{(0.2w,0.25h)}{v1}
        \fmfforce{(0.2w,0.75h)}{v2}
        \fmfforce{(0.45w,0.25h)}{v3}
        \fmfforce{(0.45w,0.75h)}{v4}
        \fmfforce{(0.65w,0.15h)}{v5}
        \fmfforce{(0.65w,0.85h)}{v6}
        \fmfforce{(0.8w,0.5h)}{v7}
        \fmf{fermion,label=$p_{1}$}{i1,v1}
        \fmf{fermion,label=$p_{2}$}{i2,v2}
        \fmf{fermion,label=$p_{3}$}{i4,v6}
        \fmf{fermion,label=$p_{4}$}{i5,v7}
        \fmf{plain}{i3,v5}
        \fmf{fermion,label=$\ell_{1}$}{v1,v3}
        \fmf{fermion,label=$\ell_{2}$}{v3,v5}
        \fmf{plain}{v2,v4,v6,v7,v5}
        \fmf{plain}{v1,v2}
        \fmf{plain}{v3,v4}
    \end{fmfgraph*}
    \caption{Massless Pentabox\label{fig:pentabox}}
\end{figure}
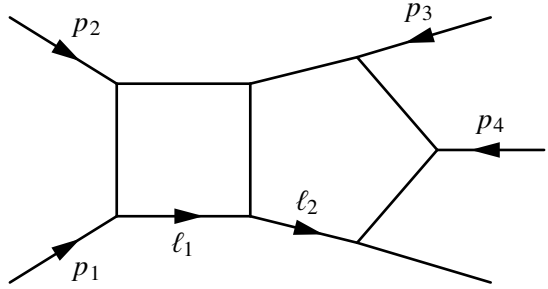

The second example is the massless pentabox, as shown in Fig. \ref{fig:pentabox}. We consider the following list of target integrals
\begin{equation}
    \{F(1,1,1,1,1,1,1,1,-10,-10,0),F(1,1,1,1,1,1,1,1,-5,-6,-3)\}\,,
\end{equation}
which go up to rank $20$. We provide statistics for the reduction of this set of integrals in Table~\ref{tab:tab3}. We attempted to reduce the less challenging rank-$14$ integral $F(1,1,1,1,1,1,1,1,-5,-6,-3)$ with Kira 3 \cite{Lange:2025fba}, and the job was killed after 23 minutes for running out of RAM whilst generating equations. (The laptop had 64 GB of total RAM, and 55.7 GB was consumed by Kira 3 when the job was killed.)
\begin{table}[h!]
\centering
    \begin{tabular}{|c|c|c|c|c|}
        \hline
        Cut&Time Taken&Number of Equations&Number of Masters\\
        \hline
        \{1,4,8\}&6997s&51619&21\\
        \hline
        \{1,5,8\}&2912s&39446&27\\
        \hline
        \{2,5,8\}&23721s&112188&31\\
        \hline
        \{1,3,4,6\}&974s&3979&13\\
        \hline
        \{1,3,4,7\}&1063s&4275&9\\
        \hline
        \{2,4,7,8\}&7064s&28338&12\\
        \hline
    \end{tabular}
    \caption{Table showing the time taken for generating the system of reduction rules at the given numerical point on a set of spanning cuts (cuts that can be obtained from these ones through symmetry relations are omitted), as well as the number of rules applied (equations) and master integrals. The sample time for applying these reduction rules using backwards substitution at a single numerical point modulo some prime number is very small compared to the times presented here.\label{tab:tab3}}
\end{table}

\subsection{Spinning Black Hole}

\begin{figure}[h!]
    \centering
    \begin{fmfgraph*}(200,100)
        \fmfforce{(0,0.2h)}{i1}
        \fmfforce{(0,0.8h)}{i2}
        \fmfforce{(w,0.2h)}{i3}
        \fmfforce{(w,0.8h)}{i4}
        \fmfforce{(0.2w,0.2h)}{v1}
        \fmfforce{(0.2w,0.8h)}{v2}
        \fmfforce{(0.5w,0.2h)}{v3}
        \fmfforce{(0.5w,0.8h)}{v4}
        \fmfforce{(0.8w,0.2h)}{v5}
        \fmfforce{(0.8w,0.8h)}{v6}
        \fmf{fermion,label=$p_{2}$}{i1,v1}
        \fmf{fermion,label=$p_{1}$}{i2,v2}
        \fmf{fermion,label=$p_{4}$}{v6,i4}
        \fmf{fermion,label=$p_{3}$}{v5,i3}
        \fmf{plain}{v1,v3,v5}
        \fmf{plain}{v2,v4,v6}
        \fmf{wiggly}{v1,v2}
        \fmf{wiggly}{v3,v6}
        \fmf{wiggly}{v5,v4}
    \end{fmfgraph*}
    \caption{Post-Minkowskian Non-Planar Double Box\label{fig:npdoublebox}}
\end{figure}
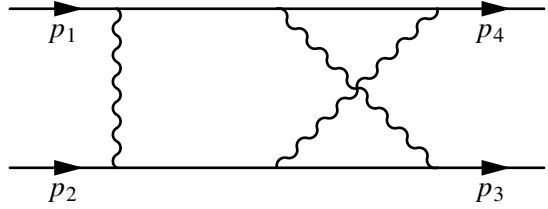

To illustrate the particular relevance of this work, we also provide a physical example. We refer directly to the quartic-in-spin, third post-Minkowskian order binary system of a spinning and spinless black hole \cite{Akpinar:2025bkt}. The most troublesome part of this calculation was the IBP reduction, for which the most complex diagram was the nonplanar double box, shown in Fig. \ref{fig:npdoublebox}. 

In this particular problem, there are $47365$ target integrals to reduce, and these can be up to rank $10$ with $10$ dots. In \cite{Akpinar:2025bkt}, the IBP reduction was completed for these integrals using a private version of FIRE which builds upon Refs.\ \cite{Smirnov:2019qkx,Smirnov:2023yhb,Smirnov:2024onl,Smirnov:2025prc} at $\sim1000$ numerical points. Each numerical point took $\sim15$ minutes to evaluate, therefore the total run time ended up being $\sim10$ days on a cluster.

Whereas the equation generation for FIRE was very fast, it took $\sim9$ hours with our algorithm to generate the system of $152113$ equations and $152104$ variables. However solving the system at each numerical point took $\sim8s$, therefore the full solving time amounts to $\sim2$ hours. Combined, this gives a total time of $\sim11$ hours.

\section{LoopIn: A framework for multi-loop calculations \label{sec:loopin}}

The IBP techniques discussed in this work will be extremely relevant for the development of \LoopIn{} ({\sc Loop In}tegrals for virtual corrections). 

\LoopIn{} v1.0 is a framework developed for the fully automated numerical evaluation of multi-loop scattering amplitudes, from amplitude generation to its numerical result. The code is built in Mathematica and interfaces several publicly available tools.
The project was developed within the research program "Advanced Calculus for Precision Physics" (ACPP), and selected as a  flagship use case for applications on HPC infrastructures, in particular on the Leonardo HPC at CINECA.

This novel code is designed to treat $2\to n$ scattering amplitudes in Standard Model and Beyond. As input, the user is required to specify the QFT model, the scattering process, the number of loops and a set of phase-space points. \LoopIn{} will then 
provide a Laurent series of the amplitude w.r.t. the dimensional regulator $\epsilon$ (where $d = 4 - 2\epsilon$), where its coefficients are numerically evaluated.\newpage

\LoopIn{} follows a modular and fully automated structure, which will carry out the following algorithmic steps:
\begin{itemize}[itemsep=0.1\baselineskip, parsep=0pt]
    
\item Amplitude generation;

\item Analysis of graph topologies;

\item Constructing interferences;

\item Reduction of target integrals:

\item Numerical evaluation of MIs;

\item Results merging and Laurent expansion.

\end{itemize}
    






\begin{figure}[h!]
    \centering
    \includegraphics[scale=0.5]{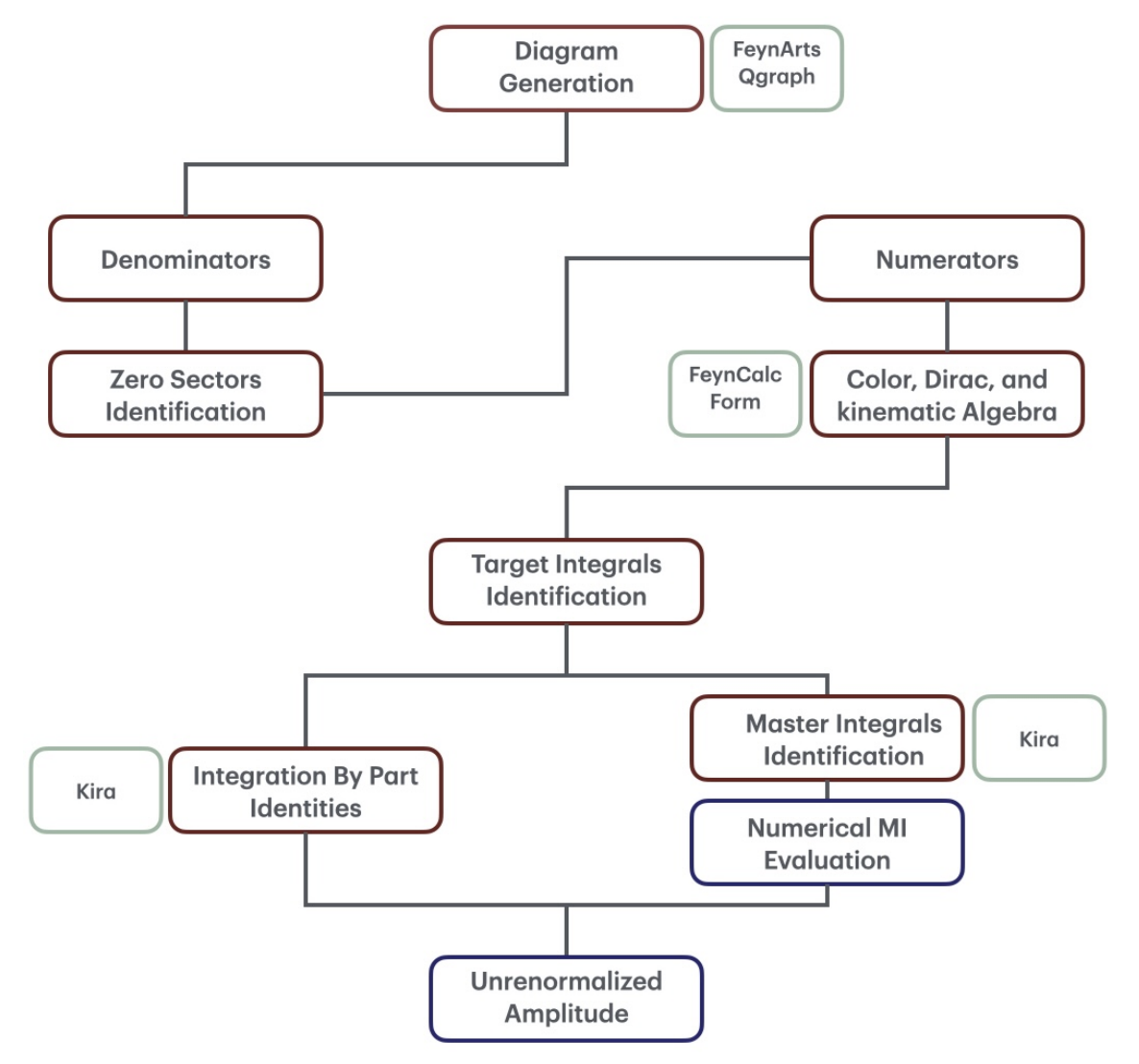}
    \caption{\LoopIn{} flowchart \label{fig:LoopInflowchart}}
\end{figure}

A visual sketch of the \LoopIn{} workflow is shown in Fig. (\ref{fig:LoopInflowchart}).

Every step is managed by a \LoopIn{} module, onto which different external and internal routines can be linked and interfaced.
The diagram generation currently uses {\sc FeynArts} \cite{Hahn:2000kx}, with an interface to {\sc QGraf} \cite{Nogueira:1991ex}
under development. The topological analysis is done by an in-house module based on the
Pak algorithm \cite{Pak:2011xt}. Graph isomorphism and automorphism techniques are under implementation,
allowing the identification of sector symmetries and crossing relations among topologies. 
\LoopIn{} then automatically generates the interferences with the tree-level amplitude. The projection of the amplitude onto form factor is currently under development as an in-house dedicated module. The numerator algebra module is linked both to {\sc FeynCalc} \cite{Shtabovenko:2016sxi} or {\sc FORM} \cite{Kuipers:2012rf}. Bash scripts are provided, allowing the parallelization of the symbolic manipulation of individual diagram interferences, together with the extrapolation of the target integrals.
The IBP reduction module is currently interfaced with {\sc Kira} \cite{Lange:2025fba}; a link with {\sc LiteRed} \cite{Lee:2012cn,Lee:2013mka} and {\sc FiniteFlow} will be available in the next update.
Lastly, the module devoted to the MIs numerical evaluation is currently interfaced with {\sc AMFlow} \cite{Liu:2022chg}; under development, additional interfaces with {\sc Line} \cite{Prisco:2025wqs} and {\sc pySecDec} \cite{Borowka:2017idc} are
under implementation. The merging of the results is currently done within \LoopIn{} with a dedicated function; a merger module is under investigation, in such a way that intermediate results can be extracted from the workflow, providing partial results.

The validation of \LoopIn{} workflow has been done on several one- and two-loop test amplitude with at most five external legs, as we report in Table \ref{tab:loopinvals}. Numerical results in individual phase-space points have been cross-checked against independent calculations, finding full agreement and confirming the robustness of \LoopIn{} as automated framework.

\begin{table}[htbp]
\centering
\begin{tabular}{lccccc}
\toprule
\textbf{Process} & \textbf{Theory} & \textbf{Loop} & \textbf{Interference Terms} & \textbf{Target Int's} & \textbf{Master Int's} \\
\midrule
$e^-\mu^+ \rightarrow e^-\mu^+$      & QED & 1 & 6    & 58       & 7   \\
\multicolumn{1}{c}{"}        & \multicolumn{1}{c}{"} & 2 & 69   & 8656     & 81  \\

$e^-\mu^+ \rightarrow e^-\mu^+\gamma$ & QED & 1 & 179  & 1032     & 54  \\
\multicolumn{1}{c}{"}        & \multicolumn{1}{c}{"} & 2 & 2688 & 220,274 & (tbd)  \\
$q\bar q \rightarrow t\bar t$          & QCD & 1 & 11   & 98       & 8   \\
\multicolumn{1}{c}{"}        & \multicolumn{1}{c}{"} & 2 & 218  & 17{,}344 & 182 \\

$gg \rightarrow t\bar t$          & QCD & 1 & 102  & 444      & 14  \\
\multicolumn{1}{c}{"}        & \multicolumn{1}{c}{"} & 2 & 1686 & 111{,}171 & 471 \\

$q\bar q \rightarrow ZZ$          & QCD+EW  & 1 & 66   & 236      & 19  \\
$gg \rightarrow HH$          & QCD+EW  & 1 & $12^*$ & 65 & 12  \\
$gg \rightarrow W^+W^-$          & QCD+EW  & 1 & $108^*$ & 420 & 16  \\
\bottomrule
\end{tabular}
\caption{Validation processes for \LoopIn{} v1.0. Asterisk entries indicates processes for which projection onto form factors have been used in replacements of interferences with the leading order amplitude.\label{tab:loopinvals}}
\label{tab:loopin_validation}
\end{table}

\section{Further Outlook \label{sec:conclusion}}

Beyond implementation within LoopIn, the algorithm presented in this work open up many possibilities for further exploration. It is clear from the spinning black hole example that the algorithm can be incredibly effective for the computation of amplitudes in non-renormalisable field theories such as gravity, where it is common to find very complex integrals. 

The main bottleneck for the current state of the algorithm is in the generation of the reduction rules. There may be some instances where standard identities without syzygy constraints fair better, as is used in LiteRed. Indeed, we have found in some cases that the use of syzygy constraints does not have a significant impact on performance time.

Furthermore, it would prove interesting to study further the technical aspects of this algorithm, such as the monomial ordering chosen, and whether there is some kind of non-commutative algebra structure present in the generated reduction rules, which has been found for IBP operators in the past (see e.g.\ \cite{Barakat:2022qlc,Barakat:2022ttc}). 

Finally, our results indicate that despite decades of work, significant improvements to the computational efficiency of IBP reduction are still possible, and intense future investigations are warranted.

\section*{Acknowledgements}

We wish to thank William J. Torres Bobadilla for organising the \textit{Scattering Amplitudes @ Liverpool} workshop earlier this year, and the participants for many interesting discussions related to this work during the workshop. We also wish to thank Giulio Crisanti for interesting discussions related to this work. 
We acknowledge the ICSC for awarding ACPP access to the EuroHPC supercomputer LEONARDO, hosted by CINECA (Italy).
S.S. research is partially supported by the Amplitudes INFN scientific initiative.
\bibliographystyle{JHEP}
\bibliography{biblio}

\end{fmffile}

\end{document}